# Degenerate solutions to the Dirac equation for massive particles and their applications in quantum tunneling

Georgios N. Tsigaridas[1,*], Aristides I. Kechriniotis[2], Christos A. Tsonos[2] and Konstantinos K. Delibasis[3]

[1]Department of Physics, School of Applied Mathematical and Physical Sciences, National Technical University of Athens, GR-15780 Zografou Athens, Greece

[2]Department of Physics, University of Thessaly, GR-35100 Lamia, Greece

[3]Department of Computer Science and Biomedical Informatics, University of Thessaly, GR-35131 Lamia, Greece

[*]Corresponding Author. E-mail: gtsig@mail.ntua.gr

**Abstract**

In a recent work we have proven the existence of degenerate solutions to the Dirac equation, corresponding to an infinite number of different electromagnetic 4-potentials and fields, providing also some examples regarding massless particles. In the present article our results are extended significantly, providing degenerate solutions to the Dirac equation for particles with arbitrary mass, which, under certain conditions, could be interpreted as pairs of particles (or antiparticles) moving in a potential barrier with energy equal to the height of the barrier and spin opposite to each other. We calculate the electromagnetic 4-potentials and fields corresponding to these solutions, providing also some examples regarding 4-potentials corresponding to both spatially constant electromagnetic fields and electromagnetic waves. Further, we discuss some potential applications of our work, mainly regarding the control of the particles outside the potential barrier, without affecting their state inside the barrier. Finally, we study the effect of small perturbations to the degenerate solutions, showing that our results are still valid, in an approximate sense, provided that the amplitude of the electromagnetic 4-potentials corresponding to the exact degenerate solutions is sufficiently small.

## 1. Introduction

In a recent article [1] we have shown that all solutions to the Dirac equation

$$i\gamma^{\mu}\partial_{\mu}\Psi + a_{\mu}\gamma^{\mu}\Psi - m\Psi = 0 \tag{1}$$

satisfying the conditions $\Psi^\dagger \gamma\, \Psi = 0$ and $\Psi^T \gamma_2\, \Psi \neq 0$ are degenerate, corresponding to an infinite number of electromagnetic 4-potentials which are explicitly calculated in Theorem 5.4. Here, $\gamma^\mu$ are the four contravariant gamma matrices in Dirac representation

$$\gamma^0 = \begin{pmatrix} 1 & 0 & 0 & 0 \\ 0 & 1 & 0 & 0 \\ 0 & 0 & -1 & 0 \\ 0 & 0 & 0 & -1 \end{pmatrix} \quad \gamma^1 = \begin{pmatrix} 0 & 0 & 0 & 1 \\ 0 & 0 & 1 & 0 \\ 0 & -1 & 0 & 0 \\ -1 & 0 & 0 & 0 \end{pmatrix}$$

$$\gamma^2 = \begin{pmatrix} 0 & 0 & 0 & -i \\ 0 & 0 & i & 0 \\ 0 & i & 0 & 0 \\ -i & 0 & 0 & 0 \end{pmatrix} \quad \gamma^3 = \begin{pmatrix} 0 & 0 & 1 & 0 \\ 0 & 0 & 0 & -1 \\ -1 & 0 & 0 & 0 \\ 0 & 1 & 0 & 0 \end{pmatrix}$$

(2)

satisfying the anticommutation relation $\{\gamma^\mu, \gamma^\nu\} = \gamma^\mu \gamma^\nu + \gamma^\nu \gamma^\mu = 2\eta^{\mu\nu} I_4$, which is essential for generating a Clifford algebra. Here, $I_4$ is the $4 \times 4$ identity matrix and $\eta^{\mu\nu}$ is the Minkowski metric with signature $(+---)$. Explicitly, $\eta^{\mu\nu} = 0$ if $\mu \neq \nu$, $\eta^{\mu\nu} = +1$ if $\mu = \nu = 0$ and $\eta^{\mu\nu} = -1$ if $\mu = \nu = 1, 2, 3$.

The matrix $\gamma$ is defined as $\gamma = \gamma^0 + i\gamma^1 \gamma^2 \gamma^3$. Further, $m$ is the mass of the particles and $a_\mu = qA_\mu$ where $q$ is the electric charge of the particles and $A_\mu$ the electromagnetic 4-potential. It should also be noted that, in order to be consistent with our previous work [1], Eq. (1) is written in natural units, where $\hbar = c = 1$.

In [1] it has also been shown that, in the case of free Dirac particles, the degenerate solutions correspond to massless particles except for particle-antiparticle pairs. However, the net charge of the particle – antiparticle pair is zero, and consequently the degeneracy is not particularly meaningful from a practical point of view. Consequently, an interesting question is the following: Are there degenerate spinors corresponding to massive particles with non-zero charge? And if yes, what is the physical interpretation of these solutions?

The first question is answered in section 2, where we provide a general class of degenerate solutions to the Dirac equation corresponding to charged particles with arbitrary mass. We also calculate the electromagnetic 4-potentials and fields corresponding to these solutions. In section 3, we discuss the physical interpretation of the degenerate solutions, showing that, under certain conditions, they could be interpreted as pairs of particles, or antiparticles, moving through a potential barrier, with energy equal to the height of the barrier and spin opposite to each other.

Further, in section 4, we provide some examples regarding the electromagnetic 4-potentials and fields corresponding to the degenerate solutions, which include 4-potentials corresponding to both spatially constant fields and electromagnetic waves.

We also discuss some potential practical applications of our results, especially regarding the control of the motion of the particles in the region of the potential barrier, without affecting their state inside the potential barrier, and consequently the transmittance through the barrier. As an example, we have calculated the combination of electric and magnetic fields that should be applied in order to accelerate the particles in the region of the potential barrier, without affecting their state inside the barrier and consequently, the transmittance through the barrier.

Here, it should be noted that, as it will be shown in section 4, in the case of degenerate solutions, the transmittance through the potential barrier for free electron pairs, becomes negligible if the width of the barrier becomes larger than $\sim 1\,pm$. Consequently, our results are expected to be applicable mainly in the framework of nuclear [2] and particle physics. However, in certain materials, as graphene [3, 4] and semiconductors [5], the effective mass of the electrons can become much smaller than their actual mass. In this case, the maximum width of the barrier permitting non-negligible transmittance becomes much larger than $1\,pm$, providing the opportunity for a wider variety of potential applications of our work, especially in fields related to nanotechnology, as nanoelectronics [6-11], graphene physics [12, 13], nanophotonics [14, 15], quantum measurements [16], etc. However, the main purpose of this article is to present of the new, interesting, physics predicted regarding the electromagnetic interactions of charged particles in the region of potential barriers. More details on the potential practical applications of our results will be provided in future works.

In section 5, we discuss the effects of small perturbations to the degenerate solutions, showing that, in an approximate sense, the particles maintain their quantum state under the influence of the electromagnetic fields corresponding to the exact degenerate solutions, provided that the magnitude of these fields is sufficiently small. Obviously, this result enhances further the potential of our results for practical applications.

Finally, in section 6, we provide a brief discussion on the physical properties of a more general class of degenerate solutions, which probably correspond to particles or pairs of particles moving in classically forbidden regions.

## 2. Mathematical form of the degenerate solutions for massive particles

In this section we will provide a general form of degenerate solutions to the Dirac equation for particles with arbitrary mass, calculating also the corresponding electromagnetic fields.

In our effort to answer the question stated in the introduction we have used a general form of degenerate spinors

$$\Psi = \begin{pmatrix} \exp(i\eta) d \sin \zeta \\ e - d \cos \zeta \\ \exp(i\eta) e \sin \zeta \\ d - e \cos \zeta \end{pmatrix} \quad (3)$$

where $d, e$ are arbitrary complex functions of the spatial coordinates and time and $\zeta, \eta$ are real constants. Using the above ansatz and requiring to be solution to the Dirac equation for real 4-potentials it is found that all spinors of the form

$$\Psi = c_1 \exp(if \cos \xi) \exp\left[-\frac{m}{\sin^2 \xi}(-z + t \cos \xi)\right] \begin{pmatrix} i \sin \xi \\ -i - \cos \xi \\ \sin \xi \\ 1 + i \cos \xi \end{pmatrix} \quad (4)$$

where $c_1$ is an arbitrary complex constant, $\xi$ a real parameter $(\xi \neq n\pi, n \in \mathbb{Z})$ and $f$ an arbitrary real function of the spatial coordinates and time, are degenerate and satisfy the Dirac equation for the 4-potentials

$$\begin{pmatrix} a_0 \\ a_1 \\ a_2 \\ a_3 \end{pmatrix} = q \begin{pmatrix} A_0 \\ A_1 \\ A_2 \\ A_3 \end{pmatrix} = \begin{pmatrix} \frac{\partial f}{\partial t} \cos \xi + \frac{\partial f}{\partial z} - g \\ \frac{\partial f}{\partial x} \cos \xi - m \cot \xi \\ \frac{\partial f}{\partial y} \cos \xi + \left(\frac{\partial f}{\partial z} - g\right) \sin \xi \\ g \cos \xi \end{pmatrix} \quad (5)$$

where $g$ is also an arbitrary real function of the spatial coordinates and time. It should be noted that, setting $g = (\partial f / \partial z)$, the above 4-potentials are simplified, taking the form

$$(a_0, a_1, a_2, a_3) = \cos \xi \left(\frac{\partial f}{\partial t}, \frac{\partial f}{\partial x} - \frac{m}{\sin \xi}, \frac{\partial f}{\partial y}, \frac{\partial f}{\partial z}\right) \quad (6)$$

Further, according to Theorem 5.4 in ref. [1], the spinors (4) will also be solutions to the Dirac equation for an infinite number of 4-potentials, given by the formula

$$b_\mu = a_\mu + s\kappa_\mu \quad (7)$$

where

$$(\kappa_0, \kappa_1, \kappa_2, \kappa_3) = \left(1, -\frac{\Psi^T \gamma^0 \gamma^1 \gamma^2 \Psi}{\Psi^T \gamma^2 \Psi}, -\frac{\Psi^T \gamma^0 \Psi}{\Psi^T \gamma^2 \Psi}, \frac{\Psi^T \gamma^0 \gamma^2 \gamma^3 \Psi}{\Psi^T \gamma^2 \Psi}\right) = (1, 0, \sin \xi, -\cos \xi) \quad (8)$$

and $s$ is an arbitrary real function of the spatial coordinates and time.

The electromagnetic fields (in Gaussian units) corresponding to the above 4-potetials are [17]

$$\mathbf{E} = -\nabla U - \frac{\partial \mathbf{A}}{\partial t} = -\frac{\partial s_q}{\partial x}\mathbf{i} + \left(\sin\xi \frac{\partial s_q}{\partial t} - \frac{\partial s_q}{\partial y}\right)\mathbf{j} - \left(\cos\xi \frac{\partial s_q}{\partial t} + \frac{\partial s_q}{\partial z}\right)\mathbf{k}$$

$$\mathbf{B} = \nabla \times \mathbf{A} = \left(\sin\xi \frac{\partial s_q}{\partial z} + \cos\xi \frac{\partial s_q}{\partial y}\right)\mathbf{i} - \cos\xi \frac{\partial s_q}{\partial x}\mathbf{j} - \sin\xi \frac{\partial s_q}{\partial x}\mathbf{k}$$

(9)

Here $U = b_0/q$ is the electric potential, $\mathbf{A} = -(1/q)(b_1\mathbf{i} + b_2\mathbf{j} + b_3\mathbf{k})$ is the magnetic vector potential, $f_q = f/q$ and $s_q = s/q$. Also, in the above equations the speed of light has been set equal to unity, since we are working in the natural system of units, where $\hbar = c = 1$.

An interesting remark is that spinors of the form (4) remain unaffected by the influence of the 4-potentials (7), regardless of the normalization constant of the spinors – parameter $c_1$ in Eq. (4) – and the amplitude of the fields. Furthermore, since the amplitude of the 4-potentials corresponding to the degenerate solutions is arbitrary, the influence of the electromagnetic fields generated by the Dirac particles (back-reaction) can be considered negligible in most cases of practical interest, and consequently, it has not been taken into account in the present work.

## 3. Physical interpretation of the degenerate solutions in the framework of quantum tunneling

In this section we shall discuss the physical interpretation of the degenerate solutions provided above, showing that, under certain conditions, they correspond to massive particles propagating in a potential barrier with energy equal to the height of the barrier and spin perpendicular to their propagation direction.

As far as the physical interpretation of the solutions (4) is considered we note that the spinor (4) can describe particles of any mass (electrons, protons, muons, etc.). However, the presence of the real exponential term $\exp\left[-m(-z + t\cos\xi)/\sin^2\xi\right]$ imposes a restriction on the coordinates $z, t$, which can only take finite values.

A similar situation occurs in the case of quantum tunneling [18], where the spinor describing the state of the particle in the region of the potential barrier includes real exponential terms of the form $\exp(\pm k\, z)$, where $k$ is given by the formula

$$k = m\left[1 - (E - V_0)^2/m^2\right]^{1/2} \tag{10}$$

with $E, V_0$ being the energy of the particle and the height of the potential barrier, respectively. It is useful to mention that the terms $\exp(\pm k\, z)$ arise from the fact that in the region of the potential barrier, where $|E-V_0|<m$, the momentum $p_z$ of the particle is imaginary, since $p_z^2 = (E-V_0)^2 - m^2 < 0$. Consequently, the momentum of the particle can be written as $p_z = \pm ik$ and the complex exponentials $\exp(\pm ip_z z)$ describing the propagation of the free particle in the form of a plane wave become the real exponential terms $\exp(\pm k\, z)$ in the region of the barrier.

An important remark is that the maximum value of the factor $k$ is equal to the mass of the particle and occurs in the case that the height of the potential barrier is equal to the energy of the particle. It is also important to note that setting the free parameter $\xi$ equal to $n\pi + \pi/2, n \in \mathbb{Z}$, the real exponential term $\exp\left[-m(-z+t\cos\xi)/\sin^2\xi\right]$ in Eq. (4) becomes $\exp(m\,z)$, and the degenerate spinors become time-independent, taking the simple form

$$\Psi = c_1 \exp(m\,z) \begin{pmatrix} \pm i \\ -i \\ \pm 1 \\ 1 \end{pmatrix} \quad (11)$$

Thus, it can be considered that the spinors given by Eq. (11) correspond to a particle of mass $m$ and energy $E$ moving along the z-axis in a potential barrier with height $V_0 = E$. This is also supported by the fact that the degenerate spinors (11) are solutions to the one-dimensional time-independent Dirac equation in the form

$$i\gamma^3 \partial_z \Psi - m\Psi = 0 \quad (12)$$

Further, in the case that $\xi = n\pi + \pi/2, n \in \mathbb{Z}$, the 4-potentials (6) become zero. Thus, the spinors (11) are also solutions to the general form of the Dirac equation (1) for zero 4-potential.

As far as the spin of the particle is concerned, we recall that the positive (when the direction of the spin is the same as the direction of motion of the particle) and negative (when the direction of the spin is opposite to the direction of motion of the particle) helicity states of a free Dirac particle (or antiparticle) are described by the following 4-vectors [19]:

$$u_\uparrow = \begin{pmatrix} \cos\left(\frac{\theta}{2}\right) \\ e^{i\varphi}\sin\left(\frac{\theta}{2}\right) \\ \frac{|\vec{p}|}{E+m}\cos\left(\frac{\theta}{2}\right) \\ \frac{|\vec{p}|}{E+m}e^{i\varphi}\sin\left(\frac{\theta}{2}\right) \end{pmatrix} \quad u_\downarrow = \begin{pmatrix} -\sin\left(\frac{\theta}{2}\right) \\ e^{i\varphi}\cos\left(\frac{\theta}{2}\right) \\ \frac{|\vec{p}|}{E+m}\sin\left(\frac{\theta}{2}\right) \\ -\frac{|\vec{p}|}{E+m}e^{i\varphi}\cos\left(\frac{\theta}{2}\right) \end{pmatrix} \quad (13)$$

$$v_\uparrow = \begin{pmatrix} \frac{|\vec{p}|}{E+m}\sin\left(\frac{\theta}{2}\right) \\ -\frac{|\vec{p}|}{E+m}e^{i\varphi}\cos\left(\frac{\theta}{2}\right) \\ -\sin\left(\frac{\theta}{2}\right) \\ e^{i\varphi}\cos\left(\frac{\theta}{2}\right) \end{pmatrix} \quad v_\downarrow = \begin{pmatrix} \frac{|\vec{p}|}{E+m}\cos\left(\frac{\theta}{2}\right) \\ \frac{|\vec{p}|}{E+m}e^{i\varphi}\sin\left(\frac{\theta}{2}\right) \\ \cos\left(\frac{\theta}{2}\right) \\ e^{i\varphi}\sin\left(\frac{\theta}{2}\right) \end{pmatrix} \quad (14)$$

where the 4-vectors $u_\uparrow, u_\downarrow$ correspond to particles and the 4-vectors $v_\uparrow, v_\downarrow$ correspond to antiparticles. Here $|\vec{p}|$ is the modulus of the momentum of the particle (or antiparticle) and $\theta, \varphi$ are the angles defining its propagation direction in spherical coordinates.

Obviously, in the case that the particle (or the antiparticle) moves along the $z-axis$, the above 4-vectors take the form

$$u_\uparrow(+z) = \begin{pmatrix} 1 \\ 0 \\ \frac{|\vec{p}|}{E+m} \\ 0 \end{pmatrix} \quad u_\downarrow(+z) = \begin{pmatrix} 0 \\ 1 \\ 0 \\ -\frac{|\vec{p}|}{E+m} \end{pmatrix} \quad u_\uparrow(-z) = \begin{pmatrix} 0 \\ 1 \\ 0 \\ \frac{|\vec{p}|}{E+m} \end{pmatrix} \quad u_\downarrow(-z) = \begin{pmatrix} -1 \\ 0 \\ \frac{|\vec{p}|}{E+m} \\ 0 \end{pmatrix} \quad (15)$$

$$v_\uparrow(+z) = \begin{pmatrix} 0 \\ -\frac{|\vec{p}|}{E+m} \\ 0 \\ 1 \end{pmatrix} \quad v_\downarrow(+z) = \begin{pmatrix} \frac{|\vec{p}|}{E+m} \\ 0 \\ 1 \\ 0 \end{pmatrix} \quad v_\uparrow(-z) = \begin{pmatrix} \frac{|\vec{p}|}{E+m} \\ 0 \\ -1 \\ 0 \end{pmatrix} \quad v_\downarrow(-z) = \begin{pmatrix} 0 \\ \frac{|\vec{p}|}{E+m} \\ 0 \\ 1 \end{pmatrix} \quad (16)$$

where the angle $\varphi$ has been set equal to zero for simplicity. In the region of the barrier $|\vec{p}| = \pm ik$, and consequently, the 4-vectors (13), (14) can be written as [18]

$$u_\uparrow(+z) = \begin{pmatrix} 1 \\ 0 \\ \dfrac{\pm ik}{|E-V_0|+m} \\ 0 \end{pmatrix} \quad u_\downarrow(+z) = \begin{pmatrix} 0 \\ 1 \\ 0 \\ \dfrac{\mp ik}{|E-V_0|+m} \end{pmatrix}$$

$$u_\uparrow(-z) = \begin{pmatrix} 0 \\ 1 \\ 0 \\ \dfrac{\pm ik}{|E-V_0|+m} \end{pmatrix} \quad u_\downarrow(-z) = \begin{pmatrix} -1 \\ 0 \\ \dfrac{\pm ik}{|E-V_0|+m} \\ 0 \end{pmatrix}$$

(17)

$$v_\uparrow(+z) = \begin{pmatrix} 0 \\ \dfrac{\mp ik}{|E-V_0|+m} \\ 0 \\ 1 \end{pmatrix} \quad v_\downarrow(+z) = \begin{pmatrix} \dfrac{\pm ik}{|E-V_0|+m} \\ 0 \\ 1 \\ 0 \end{pmatrix}$$

$$v_\uparrow(-z) = \begin{pmatrix} \dfrac{\pm ik}{|E-V_0|+m} \\ 0 \\ -1 \\ 0 \end{pmatrix} \quad v_\downarrow(-z) = \begin{pmatrix} 0 \\ \dfrac{\pm ik}{|E-V_0|+m} \\ 0 \\ 1 \end{pmatrix}$$

(18)

In the case that $E = V_0$, $k = m$, and the 4-vectors (17), (18) take the simpler form

$$u_\uparrow(+z) = \begin{pmatrix} 1 \\ 0 \\ \pm i \\ 0 \end{pmatrix} \quad u_\downarrow(+z) = \begin{pmatrix} 0 \\ 1 \\ 0 \\ \mp i \end{pmatrix} \quad u_\uparrow(-z) = \begin{pmatrix} 0 \\ 1 \\ 0 \\ \pm i \end{pmatrix} \quad u_\downarrow(-z) = \begin{pmatrix} -1 \\ 0 \\ \pm i \\ 0 \end{pmatrix} \quad (19)$$

$$v_\uparrow(+z) = \begin{pmatrix} 0 \\ \mp i \\ 0 \\ 1 \end{pmatrix} \quad v_\downarrow(+z) = \begin{pmatrix} \pm i \\ 0 \\ 1 \\ 0 \end{pmatrix} \quad v_\uparrow(-z) = \begin{pmatrix} \pm i \\ 0 \\ -1 \\ 0 \end{pmatrix} \quad v_\downarrow(-z) = \begin{pmatrix} 0 \\ \pm i \\ 0 \\ 1 \end{pmatrix} \quad (20)$$

Assuming that the contributions of the positive and negative helicity states are equal, the 4-vector describing the spin of the particle (or antiparticle) in the region of the barrier become

$$u_{\uparrow\downarrow}(+z) = \begin{pmatrix} 1 \\ 1 \\ \pm i \\ \mp i \end{pmatrix} \qquad u_{\uparrow\downarrow}(-z) = \begin{pmatrix} -1 \\ 1 \\ \pm i \\ \pm i \end{pmatrix} \tag{21}$$

$$v_{\uparrow\downarrow}(+z) = \begin{pmatrix} \pm i \\ \mp i \\ 1 \\ 1 \end{pmatrix} \qquad v_{\uparrow\downarrow}(-z) = \begin{pmatrix} \pm i \\ \pm i \\ -1 \\ 1 \end{pmatrix} \tag{22}$$

Based on the above analysis, it can be assumed that the degenerate spinor (11) corresponds to a pair of particles (or antiparticles) with spin opposite to each other, in order to ensure equal contribution of the positive and negative helicity eigenstates. It should also be mentioned that this interpretation can be applied (equally well) to both massive and massless particles.

In general, choosing the upper sign in the above 4-vectors, it can be easily verified that the spinors

$$\Psi_p = c_+ \exp(-m\,z) \begin{pmatrix} 1 \\ 1 \\ i \\ -i \end{pmatrix} + c_- \exp(m\,z) \begin{pmatrix} -1 \\ 1 \\ i \\ i \end{pmatrix} \tag{23}$$

$$\Psi_a = c_+ \exp(m\,z) \begin{pmatrix} i \\ -i \\ 1 \\ 1 \end{pmatrix} + c_- \exp(-m\,z) \begin{pmatrix} i \\ i \\ -1 \\ 1 \end{pmatrix} \tag{24}$$

where $c_+$, $c_-$ are arbitrary complex constants corresponding to motion along the +z and -z direction respectively, are degenerate solutions to the one-dimensional time-independent Dirac equation (12) and consequently to the general form of the Dirac equation (1) for zero 4-potential. These solutions can be interpreted as pairs of particles (23) or antiparticles (24) moving along the $\pm z$ direction in a potential barrier with height equal to the energy of the particles (or antiparticles). Also, the spins of the particles (or antiparticles) are opposite to each other, in order to ensure equal contribution of the positive and negative helicity eigenstates. A schematic representation of the quantum tunneling for a pair of Dirac fermions in the case of degenerate solutions is shown in Figure 1.

Further, since the choice of the coordinate system is arbitrary, the z-axis can be set to correspond to any desired direction in space. Thus, Eqs. (23), (24) describe a pair of particles (or antiparticles) with spin opposite to each other, moving along the z-axis which can be set to correspond to any desired direction in space, in a potential barrier with height equal to the energy of the particles.

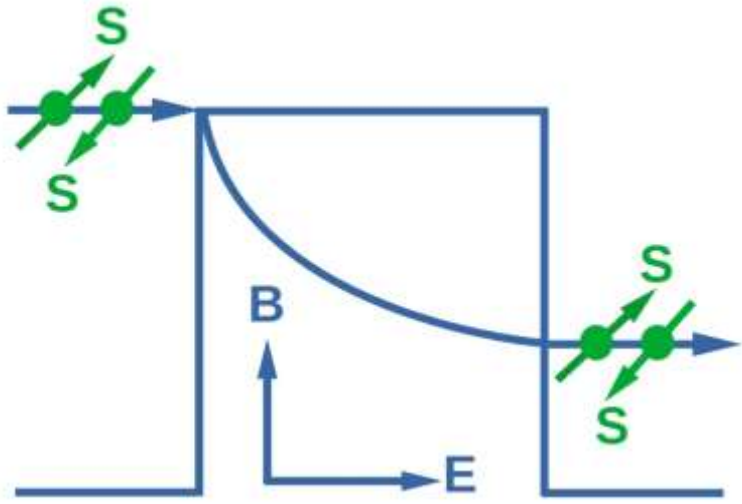

**Figure 1**: Quantum tunneling of a pair of Dirac particles in the case of degenerate solutions.

Further, in this case, the 4-potentials (7) become

$$(b_0, b_1, b_2, b_3) = s(1, 0, 1, 0) \tag{25}$$

corresponding to the following electromagnetic fields

$$\mathbf{E} = -\frac{\partial s_q}{\partial x}\mathbf{i} + \left(\frac{\partial s_q}{\partial t} - \frac{\partial s_q}{\partial y}\right)\mathbf{j} - \frac{\partial s_q}{\partial z}\mathbf{k}, \qquad \mathbf{B} = \frac{\partial s_q}{\partial z}\mathbf{i} - \frac{\partial s_q}{\partial x}\mathbf{k} \tag{26}$$

The above fields can also be considered as a special case of the electromagnetic fields (9) corresponding to $\cos\xi = 0$ and $\sin\xi = 1$.

Thus, a pair of particles (or antiparticles) with spin opposite to each other, moving along the $\pm z$ direction in a potential barrier with height equal to the energy of the particles (or antiparticles), will be in the same state either in the case that the 4-potential is zero or in the presence of any of the 4-potentials of the form (7).

Interestingly, there is also another family of degenerate solutions, corresponding to the lower sign of the 4-vectors (21), (22). Specifically, the spinors

$$\Psi'_p = c_+ \exp(m\,z)\begin{pmatrix} 1 \\ 1 \\ -i \\ i \end{pmatrix} + c_- \exp(-m\,z)\begin{pmatrix} -1 \\ 1 \\ -i \\ -i \end{pmatrix} \tag{27}$$

$$\Psi'_a = c_+ \exp(-m\,z) \begin{pmatrix} -i \\ i \\ 1 \\ 1 \end{pmatrix} + c_- \exp(m\,z) \begin{pmatrix} -i \\ -i \\ -1 \\ 1 \end{pmatrix} \tag{28}$$

are degenerate solutions to the one-dimensional time-independent Dirac equation (12) and consequently to the general form of the Dirac equation (1) for zero 4-potential. The physical interpretation of these solutions is the same as in the case of the spinors (23), (24).

In this case, the 4-potentials (7) become

$$(b_0, b_1, b_2, b_3) = s(1, 0, -1, 0) \tag{29}$$

corresponding to the electromagnetic fields

$$\mathbf{E}'(\mathbf{r},t) = -\frac{\partial s_q}{\partial x}\mathbf{i} - \left(\frac{\partial s_q}{\partial t} + \frac{\partial s_q}{\partial y}\right)\mathbf{j} - \frac{\partial s_q}{\partial z}\mathbf{k}, \qquad \mathbf{B}'(\mathbf{r},t) = -\frac{\partial s_q}{\partial z}\mathbf{i} + \frac{\partial s_q}{\partial x}\mathbf{k} \tag{30}$$

which also result from the electromagnetic fields (9) setting $\cos\xi = 0$ and $\sin\xi = -1$.

Consequently, in the case that $\cos\xi = 0$, the solutions (4) can be interpreted as a pair of particles (or antiparticles) with spin opposite to each other, moving along the z-axis in a potential barrier with height equal to the energy of the particles. In the case that $\cos\xi \neq 0$, the physical properties of the solutions (4) will be discussed in section 6.

### 4. A discussion on the potential practical applications of the above results

In this section we shall discuss some potential practical applications of our results, especially regarding the control of the state of the particles in the region of the potential barrier, without affecting their state inside the barrier, and consequently the transmittance through the potential barrier.

As far as the practical applications of the degeneracy are concerned, an obvious one is that Eq. (25) describes the set of 4-potentials not affecting the state of the particles inside the potential barrier, and consequently the transmittance through the barrier, in the case that the spinor of the particles or antiparticles are given by Eqs. (23), (24) respectively. Similarly, Eq. (29) describes the set of 4-potentials not affecting the state of the particles inside the potential barrier, and consequently the transmittance through the barrier, in the case that the spinor of the particles or antiparticles are given by Eqs. (27), (28) respectively. For example, supposing that the arbitrary function $s_q$ depends only on time, Eqs. (25), (29) imply that the state of the particles will not be affected by the presence of 4-potentials corresponding to a spatially constant, but with arbitrary time-dependence, electric field applied along the y-axis. Further, since the choice of the coordinate system is arbitrary, the y-axis can be set to correspond to any desired direction in space, perpendicular to the direction of motion of the particles. Thus, Eqs. (25), (29) imply that the state of the particles inside the potential barrier, and consequently the transmittance through the barrier, will not change in the presence of 4-potentials corresponding to a spatially constant, but with arbitrary time-dependence, electric field applied in a direction perpendicular to the direction of motion of the particles.

Further, setting

$$s_q = -E_{W1}\cos\left[k_W(y+t)+\delta_{W1}\right]x - E_{W2}\left[k_W(y+t)+\delta_{W2}\right]z \tag{31}$$

in Eqs. (26), the resulting electromagnetic fields are

$$\mathbf{E}_W = E_{W1}\cos\left[k_W(y+t)+\delta_{W1}\right]\mathbf{i} + E_{W2}\cos\left[k_W(y+t)+\delta_{W2}\right]\mathbf{k} \tag{32}$$

$$\mathbf{B}_W = -E_{W2}\cos\left[k_W(y+t)+\delta_{W2}\right]\mathbf{i} + E_{W1}\cos\left[k_W(y+t)+\delta_{W1}\right]\mathbf{k} \tag{33}$$

corresponding to a plane electromagnetic wave, of arbitrary polarization, propagating along the $-y$ direction, with Poynting vector

$$\begin{aligned}\mathbf{S} &= \frac{1}{4\pi}\mathbf{E}_W \times \mathbf{B}_W \\ &= -\frac{1}{4\pi}\left[E_{W1}^{\ 2}\cos^2\left[k_W(y+t)+\delta_{W1}\right]+E_{W2}^{\ 2}\cos^2\left[k_W(y+t)+\delta_{W2}\right]\right]\mathbf{j}\end{aligned} \tag{34}$$

In the above equations, $E_{W1},\delta_{W1},E_{W2},\delta_{W2}$ are arbitrary real constants corresponding to the amplitude and phase of the $-x$ and $-z$ component of the electric field of the wave respectively, and $k_W$ is a real parameter corresponding to the wavenumber. Further, since the choice of the coordinate system is arbitrary, the y-direction can be set to correspond to any desired direction in space, perpendicular to the direction of motion of the particles. Thus, the state of the particles inside the potential barrier, and consequently the transmittance though the barrier, will not change in the presence of 4-potentials corresponding to a plane electromagnetic wave, e. g. a laser beam, with arbitrary polarization, propagating along a direction perpendicular to the direction of motion of the particles. It can be easily shown that the same is also true for particles (or antiparticles) described by spinors of the form (27), (28).

As far as the length scale of the tunneling effect is concerned, it should be mentioned that, in Gaussian units, the real exponential terms in the time-independent degenerate spinors given by Eqs. (23), (24) or (27), (28) take the form

$$\exp\left(\pm\frac{mc}{\hbar}z\right) = \exp\left(\pm\frac{z}{z_0}\right) \tag{35}$$

where

$$z_0 = \frac{\hbar}{mc} \tag{36}$$

Thus, the parameter $z_0$ can be considered as a measure of the length scale of the tunneling effect in the case of degenerate spinors. For example, assuming a pair of electrons or positrons $\left(m_e = 9.109 \times 10^{-28}\,\text{g}\right)$, $z_0$ takes the value $z_0 = 1.93 \times 10^{-11}\,\text{cm}$ which is obviously in the subatomic level.

It is also interesting to examine the behavior of the transmittance through the potential barrier in the case of degenerate solutions. Based on Eq. (6) in [20], it is easy to calculate the transmission coefficient in the case of degenerate solutions $(V_0 = E)$, given by the formula

$$T_C = \frac{\exp(-i\,pl/\hbar)}{\cosh(l/z_0) + i(mc/p)\sinh(l/z_0)} \tag{37}$$

where $p = \sqrt{(E/c)^2 - (mc)^2}$ is the modulus of the momentum of the particle outside the potential barrier and $l$ is the width of the barrier. Thus, the transmittance through the barrier in the case of degenerate solutions is

$$T = |T_C|^2 = \frac{1}{\cosh^2(l/z_0) + (mc/p)^2 \sinh^2(l/z_0)} \tag{38}$$

From the above formula it is clear that the transmittance takes small values for slow moving particles $(p \ll mc)$, tending to zero as $p \to 0$. On the other hand, in the extreme relativistic limit $(p \gg mc)$, the transmittance takes its maximum value

$$T_{\max} = \text{sech}^2(l/z_0) \tag{39}$$

Obviously, in the case of degenerate solutions, the transmittance through the barrier does not change in the presence of the 4-potentials (25), for particles described by spinors of the form (23), (24) or (29), for particles described by spinors of the form (27), (28).

Another important remark is that, according to Eqs. (38), (39), the transmittance through the potential barrier becomes negligible, if the width of the barrier becomes larger than 2-3 times $z_0$. In the case of free electrons $z_0 = 1.93 \times 10^{-13}\,\text{m}$, and consequently our theory is mainly applicable in the framework of nuclear and particle physics. However, as it is also mentioned in the introduction, in several materials of practical interest, as graphene and semiconductors, the effective mass of the electrons can take much smaller values than the actual mass of the electron. In this case, the value of $z_0$ - and consequently the maximum width of the barrier for non-negligible transmittance - increases significantly, providing the opportunity to extend the applicability of our results to other fields, as nanoelectronics, graphene physics, nanophotonics, quantum measurements, etc.

Let us assume that an electric field $\mathbf{E} = E_0 \mathbf{k}$ is applied along the z-axis in order to accelerate a pair of particles moving along the same axis in free space. It is also assumed that the state of the particles inside the potential barrier is described by a degenerate spinor of the form (23), (24). Then, the electric potential becomes $U = -E_0 z$, and according to Eq. (25) the magnetic vector potential should become $\mathbf{A} = (0, E_0 z, 0)$ in order to preserve the state of the particles inside the barrier. This leads to a constant magnetic field $\mathbf{B} = -E_0 \mathbf{i}$ along the x-axis. Thus, a magnetic force $\mathbf{F}_B = (q/c)\mathbf{v} \times \mathbf{B} = -(q v_0 E_0 / c)\mathbf{j}$ is exerted to the particles, altering their direction of motion. Here, $\mathbf{v} = v_0 \mathbf{k}$ is the velocity of the particles in the region of the potential barrier (outside the barrier).

In order to avoid this side-effect, a constant electric field $\mathbf{E}_y = (v_0 E_0 / c)\mathbf{j}$ should be applied along the y-axis, in order to cancel out the magnetic force. In this case, the electric potential becomes $U' = -E_0(z + v_0 y / c)$. Consequently, the magnetic vector potential should also take the form $\mathbf{A}' = (0, E_0(z + v_0 y / c), 0)$ in order to preserve that state of the particles inside the barrier. However, the magnetic field, and consequently the magnetic force exerted on the particles is the same for both vector potentials $\mathbf{A}'$, $\mathbf{A}$. Therefore, the total electromagnetic force exerted on the particles is $\mathbf{F} = q\mathbf{E} + (q/c)\mathbf{v} \times \mathbf{B} = qE_0 \mathbf{k}$, the same as in the case of a constant electric field along the z-axis. Thus, the combination of the electromagnetic fields $\mathbf{E} = (v_0 E_0 / c)\mathbf{j} + E_0 \mathbf{k}$, $\mathbf{B} = -E_0 \mathbf{i}$ ensures that both the direction of motion of the particles in free space and their state inside the potential barrier are preserved. Further, since the choice of the coordinate system is arbitrary, the axes -x, -y can be set to any desired axes in space, perpendicular to each other and to the direction of motion of the particles.

Thus, the above analysis implies that, if a constant electric field with modulus $E_0$ is applied along the direction of motion of a pair of particles in order to accelerate them, then both the state of the particles inside the potential barrier and their direction of motion outside the barrier will not be affected if an electric field with modulus $(v_0 E_0 / c)$ and a magnetic field with modulus $E_0$ are applied in addition to the initial electric field, as far as they are perpendicular to each other, and the direction of their cross product $(\mathbf{E} \times \mathbf{B})$ is the same as the direction of motion of the particle.

It should be noted that, in the above analysis, the velocity of the particles in the vicinity of the barrier has been considered constant. This is justified by the fact that, as it has been shown in the analysis regarding the length scale of the tunneling, the width of the barrier should be very small, of the order of a few pm or less, in order to have non-negligible transmittance. Consequently, it is a reasonable to assume that the velocity of the particles is constant in such a small area of space.

Thus, it is clear that the property of particles described by degenerate spinors to be in the same state under a wide variety of electromagnetic 4-potentials and fields, provides the opportunity to apply certain combinations of 4-potentials in order to manipulate the motion of the particles in free space, without affecting their state inside the potential barrier, and consequently, the transmittance through the barrier.

### 5. Nearly degenerate solutions

In this section we shall study the effect of perturbations to the degenerate solutions, showing that our results are still valid, provided that the magnitude of the involved electromagnetic fields is sufficiently small.

In the above analysis it has been assumed that the solutions are degenerate, namely the spin of the particles are opposite to each other and the energy of the particles is equal to the height of the potential barrier. The first condition can be easily met in practice, since, in the case of electrons, the projection of the spin on their direction of motion can only take two values, $\pm 1/2$ in natural units, and consequently it is always easy to select two electrons with opposite values of the projection of their spin. However, the second condition is stricter, since it is not always easy to set the energy of the particles exactly equal to the height of the barrier. Therefore, it would be interesting to examine what happens in the case of small deviations from this condition.

For this purpose, we consider a pair of particles with opposite spin moving along the +z-direction in the region of a potential barrier with height $V_0 \neq E$. According to Eq. (15), the spinor corresponding to this pair of particles can be written as

$$\Psi_p(+z) = \exp(-k\ z) \begin{pmatrix} 1 \\ 1 \\ \dfrac{ik}{|E-V_0|+m} \\ \dfrac{-ik}{|E-V_0|+m} \end{pmatrix} \qquad (40)$$

where $k$ is given by Eq. (10), namely $k = m\left[1-\left(E-V_0\right)^2/m^2\right]^{1/2}$. Assuming that we are close to the degeneracy condition, namely $|E-V_0|/m = e \ll 1$, the spinors (40) can be written in the nearly degenerate form

$$\Psi_{d,p}(+z) = c_0 \exp(-k\,z) \begin{pmatrix} 1 \\ 1 \\ \dfrac{i\,k/m}{1+e} \\ \dfrac{-i\,k/m}{1+e} \end{pmatrix} \approx c_0 \exp(-m\,z) \begin{pmatrix} 1 \\ 1 \\ i(1-e) \\ -i(1-e) \end{pmatrix} \tag{41}$$

Substituting the above spinors into the Dirac equation (1) for zero 4-potential, we obtain that

$$i\gamma^\mu \partial_\mu \Psi_{d,p}(+z) - m\Psi_{d,p}(+z) = c_0 e m \exp(-m\,z) \begin{pmatrix} -1 \\ -1 \\ i \\ -i \end{pmatrix} \tag{42}$$

The right-hand term in Eq. (42) can be ignored, provided that $em \ll m$, or $e \ll 1$. Consequently, in the limit that $e \ll 1$ the perturbed spinors (41) can be considered approximate solutions to the Dirac equation for zero 4-potential. The physical interpretation of the condition $e \ll 1$ is that the deviation of the energy of the particles from the height of the potential barrier must be much smaller than the rest energy of the particles.

In their unperturbed form $(e=0)$, the spinors (41) are also solutions to the Dirac equation for the 4-potentials given by Eq. (25), namely $(b_0, b_1, b_2, b_3) = s(1,0,1,0)$, where $s$ is an arbitrary real function of the spatial coordinates and time. However, in the case that $e \neq 0$, the Dirac equation for the spinors (41) and the 4-potentials (25) takes the form

$$i\gamma^\mu \partial_\mu \Psi_{d,p}(+z) + b_\mu \gamma^\mu \Psi_{d,p} - m\Psi_{d,p}(+z)$$
$$= c_0 e m \exp(-m\,z) \begin{pmatrix} -1 \\ -1 \\ i \\ -i \end{pmatrix} + c_0 e s \exp(-m\,z) \begin{pmatrix} 1 \\ 1 \\ i \\ -i \end{pmatrix} \tag{43}$$

The right-hand terms in Eq. (43) can be ignored, if $em \ll m$ and $e|s| \ll m$. Consequently, the spinors (41) can also be considered as approximate solutions to the Dirac equation (1) for the 4-potentials (25), provided that $e \ll 1$ and $e|s| \ll m$. The physical interpretation of the condition $e|s| \ll m$ is that the energy $|s| = q|s_q|$ corresponding to the electromagnetic potentials induced by the function $s$ must be much smaller than the rest energy of the particles. Further, the larger the deviation from the exact degenerate solution is, the smaller the values of the function $s$ should be. Obviously, since the choice of the coordinate system is arbitrary, the +z - direction can be set to correspond to any desired direction in space, and consequently, the above analysis can also be applied to particles (or antiparticles) moving along any direction in space, in the region of a potential barrier.

Thus, the results obtained in this article regarding the properties of degenerate solutions will still be valid in the case of small perturbations to these solutions, provided that the perturbations are sufficiently small and the electromagnetic 4-potentials and fields corresponding to the exact degenerate solutions are sufficiently weak, in order to satisfy the conditions $e \ll 1$ and $e|s| \ll m$ in natural units.

## 6. A brief discussion on the physical properties of the general form of the degenerate solutions

As far as the general form (4) of the degenerate spinors is concerned, it should be noted that, in Gaussian units, the real exponential term $\exp\left[-\dfrac{m}{\sin^2 \xi}\left(-z + t\cos\xi\right)\right]$ becomes

$$\exp\left[-\dfrac{mc}{\hbar \sin^2 \xi}\left(-z + ct\cos\xi\right)\right] \tag{44}$$

Thus, the degenerate spinor (4) decays with a rate of $\left(mc^2/\hbar\right)\left(\cos\xi/\sin^2 \xi\right)$ as far as its temporal component is concerned and $\left(mc/\hbar\right)\left(1/\sin^2 \xi\right)$ regarding its spatial component. In the case of a pair of free electrons (or positrons), the above decay rates take the numerical values $1.554 \times 10^{21}\left(\cos\xi/\sin^2 \xi\right) s^{-1}$ for the temporal component and $5.181 \times 10^{10}\left(1/\sin^2 \xi\right) cm^{-1}$ for the spatial one. Obviously, in the special case that $\cos\xi = 0$, we obtain the time-independent solutions (11), and more general (23), (24) or (27), (28) which can be interpreted as pairs of particles, with opposite spins, impinging on a potential barrier with energy equal to the height of the barrier. The physical properties of these solutions are thoroughly discussed in section 4.

An interesting aspect of the degenerate solutions (4) is that, contrary to the special case (4), they exhibit a very rapid temporal decay, except if the parameter $\xi$ is chosen extremely close to $\pi/2$, or $n\pi + \pi/2, n \in \mathbb{Z}$ in general. However, in this case, the general solutions (4) can be considered as a perturbed form of the steady state solution (11). Thus, we could interpret the general solutions (4) as particles, or pairs of particles, moving in a classically forbidden region. Further, as the mass of the particles reduces, both the temporal and spatial decay rate become lower, tending to zero as the particles become massless. This result is expected from a physical point of view, because, in the case of massless particles, the limitations of conventional physics are removed, enabling them to move unrestricted in classically forbidden regions, where the degeneracy discussed in this article, and perhaps other interesting phenomena, occur. The case of massless particles is studied in detail in [21].

However, in this article, we have shown that even in the case of massive particles, new, extraordinary phenomena can take place regarding their electromagnetic interactions, which become more prominent as the effective mass of the particles tends to zero, which is plausible in the framework of solid-state physics.

Obviously, further extensions to our work are also possible. For example, it would be interesting to investigate the connection of our results to quantum field theory [22, 23] and general relativity [23, 24]. However, this task is quite complex and demanding and it is not possible to be covered in the present article. Hopefully, it will be investigated in a future work.

Nevertheless, we believe that the results already presented in this article are particularly interesting and ground-breaking, having the potential to trigger a plethora of further research and developments, both in theoretical and applied physics.

## 7. Conclusions

In conclusion, we have found degenerate solutions to the Dirac equation for particles with arbitrary mass, which, under certain conditions, could be interpreted as pairs of particles (or antiparticles) with spin opposite to each other, moving in a potential barrier with energy equal to the height of the barrier. We have calculated the electromagnetic 4-potentials and fields corresponding to these solutions and we have also discussed some practical applications, mainly regarding the control of the quantum state of the particles outside the potential barrier, without affecting their state inside the barrier, and consequently the transmittance through the barrier. We have also examined the effect of perturbations on the degenerate solutions, and found that, in an approximate sense, the quantum state of the particles still remains unaltered under the influence of the electromagnetic fields corresponding to the exact degenerate solutions, provided that the magnitude of these fields is sufficiently small. Finally, we have discussed the physical properties of a more general class of degenerate solutions, which probably correspond to particles moving in classically forbidden regions.